\documentclass[conference,compsoc]{IEEEtran}

\usepackage{graphicx}
\usepackage{float}
\usepackage{subcaption}

\usepackage{booktabs}
\usepackage{multirow}

\usepackage{csquotes}

\usepackage{xspace}

\usepackage[numbers]{natbib}

\usepackage{hyperref}

\usepackage{orcidlink}

\usepackage[nolist]{acronym}
\begin{acronym}
    \acro{ASIC}{Application Specific Integrated Circuit}
    \acro{CSR}{Control and Status Register}
    \acro{DRC}{Design Rule Check}
    \acro{EDA}{Electronic Design Automation}
    \acro{EM}{electro-magnetic}
    \acro{GDS}{Graphic Design System}
    \acro{IC}{Integrated Circuit}
    \acro{IO}[I/O]{Input/Output}
    \acro{IP}{Intellectual Property}
    \acro{LVS}{Layout versus Schematic}
    \acro{ML}{Machine Learning}
    \acro{OS}{Operating System}
    \acro{PDK}{Process Design Kit}
    \acro{PMP}{Physical Memory Protection}
    \acro{SEM}{Scanning Electron Microscope}
    \acro{SoC}{System-on-Chip}
    \acroplural{SoC}[SoCs]{Systems-on-Chip}
    \acro{STI}{Shallow Trench Isolation}
\end{acronym}

\newcommand{\ie}{i.\,e.}
\newcommand{\eg}{e.\,g.}
\newcommand{\etal}{et al.\@\xspace}

\begin{document}

\title{Hardware Trojans from Invisible Inversions:\\On the Trojanizability of Standard Cell Libraries}


\author{
\IEEEauthorblockN{Kolja Dorschel\IEEEauthorrefmark{1} \orcidlink{0009-0004-8593-2866}, 
René Walendy\IEEEauthorrefmark{1} \orcidlink{0000-0002-5378-3833}, 
Lukas Plätz\IEEEauthorrefmark{2} \orcidlink{0009-0006-0253-0883}, \\
Thorben Moos\IEEEauthorrefmark{3} \orcidlink{0000-0003-3809-9803}, 
Christof Paar\IEEEauthorrefmark{1} \orcidlink{0000-0001-8681-2277}, 
and Steffen Becker\IEEEauthorrefmark{2}\IEEEauthorrefmark{1} \orcidlink{0000-0001-7526-5597}}
\IEEEauthorblockA{\IEEEauthorrefmark{1}Max Planck Institute for Security and Privacy, \{firstname.lastname\}@mpi-sp.org}
\IEEEauthorblockA{\IEEEauthorrefmark{2}Ruhr University Bochum, \{firstname.lastname\}@rub.de}
\IEEEauthorblockA{\IEEEauthorrefmark{3}UC Louvain, \{firstname.lastname\}@uclouvain.be}
}

\maketitle

\begin{abstract}
\noindent
At S\&P 2023, Puschner~\etal made a valuable dataset for hardware Trojan detection research publicly available.
It contains a complete set of \acf{SEM} images of four different digital \acp{IC} fabricated at progressively smaller semiconductor technology nodes.
Puschner~\etal reported preliminary evidence that feature sizes affect Trojan detection performance, but they were unable to disentangle effects caused by insertion strategies or by degrading image quality from those intrinsic to the underlying standard cell libraries.
Distinguishing those causes, however, is crucial to understand whether improved tooling (\eg, higher resolution imaging equipment) can remove the observed technology bias, or whether susceptibility to stealthy hardware Trojans is indeed an inherent property of a cell library.
In this work, we dive deep into the S\&P 2023 dataset to answer these questions.
We devise alternative metrics to those of Puschner~\etal, in order to assess and compare the potential susceptibility of standard cell libraries more meaningfully.
We find clear differences between the evaluated process nodes.
However, in all cases we identify cells that implement distinct logic functions yet are visually indistinguishable in backside \ac{SEM} images.
We exploit this property to construct stealthy, standard-cell-based hardware Trojans and present a concrete case study: a privilege-escalation backdoor in an Ibex RISC-V core.
Our results demonstrate that cell libraries can -- and \emph{should} -- be evaluated for their potential \enquote{Trojanizability}, and we recommend practical defenses.
\end{abstract}


\IEEEpeerreviewmaketitle

\section{Introduction}
\label{sec:introduction}
Silicon-level hardware Trojans pose a severe threat to the security and integrity of modern information systems~\cite{DBLP:journals/computer/KarriRRT10}.
At the transistor and layout level, an adversary can introduce tiny modifications to an \acf{IC} that alter its logical behavior~\cite{DBLP:conf/ches/BeckerRPB13}, degrade its reliability~\cite{MOSSA2017116}, or create covert channels for data exfiltration or privilege escalation~\cite{DBLP:conf/asiacrypt/EnderG0P17,DBLP:conf/iscas/PerezIVP21,DBLP:conf/sp/YangHDAS16}.
Unlike software vulnerabilities, such modifications cannot be patched or removed once fabricated and may remain effective throughout the device's operational lifetime.
They are difficult to detect by functional testing~\cite{DBLP:conf/ccs/GohilGPR22}, and, when carefully crafted, can be made extremely small and physically localized~\cite{DBLP:conf/sp/YangHDAS16} so as to evade conventional optical inspection and electrical testing~\cite{DBLP:conf/ches/BeckerRPB13}.
Because \acp{IC} form the trusted foundation of virtually every modern computing platform, including data centers, mobile devices and embedded controllers in critical infrastructure, silicon Trojans have the potential to undermine confidentiality, integrity, and availability across a broad range of applications.

A central factor that amplifies this threat is the structure of the modern semiconductor supply chain~\cite{DBLP:journals/computer/KarriRRT10}.
Most design houses that develop \ac{IP} cores, \acp{ASIC} or \acp{SoC} do not own or operate state-of-the-art fabrication facilities anymore~\cite{Kleinhans2020GlobalSemiconductor}.
Instead, the economics of semiconductor manufacturing led to a division of labor, as design houses create digital layouts specifying transistor geometries and interconnects (typically in form of \ac{GDS}~II files), while third-party foundries operate the specialized fabrication processes required to realize those designs in silicon~\cite{DBLP:journals/pieee/RajendranSK14}.
This outsourcing model is widespread and pragmatic.
Foundries offer advanced process nodes by making large capital investments, operating relatively expensive equipment and maintaining process know-how that would be infeasible for individual design firms to replicate.
However, the separation of design and fabrication creates supply chain trust dependencies~\cite{DBLP:journals/pieee/RajendranSK14,Rekhi2025Collusion}.
Design files must be transferred securely to the foundry, mask sets and wafers must be handled and transported without tampering, and the manufacturing process must implement the intended layout and process steps exactly according to specification.

These dependencies create multiple adversarial opportunities.
In-transit or insider modification of layout files or mask sets can introduce malicious changes that are functionally subtle yet security-significant~\cite{9956883}.
An untrusted or compromised foundry, a rogue process engineer, or an interceptor during file transfer can insert, replace, or alter logic cells in ways that are difficult to discover after fabrication.
The hardware nature of such attacks means that even rigorous software-level mitigation and post-deployment updates cannot necessarily correct or fully detect the fault, particularly when Trojans are designed to remain dormant until triggered by specific conditions.

To address these risks, the semiconductor and security communities have developed and adopted physical inspection workflows intended to provide evidence that fabricated devices conform to their intended digital designs.
Inspection approaches span a spectrum from fully non-destructive imaging of each device to destructive, high-resolution analysis of a randomly sampled small representative subset.
Non-destructive approaches, such as X-ray tomography and infrared imaging, theoretically allow per-sample verification without sacrificial processing but so far either require particle beams from a synchrotron as a light source~\cite{PMID:28300088,Holler2019Threedimensional} or are limited in the spatial resolution they may provide~\cite{DBLP:journals/corr/abs-2303-07406}.
More realistic high-resolution techniques that do not require access to a particle accelerator typically involve mechanical milling and/or chemical etching followed by \acf{SEM} imaging. These are destructive but can reveal nanoscale layout structures and routing.
Yet, they cannot be applied for exhaustive verification because they typically permanently destroy inspected samples and also require expensive equipment and long operator hours~\cite{DBLP:conf/sp/PuschnerMBKMP23}.

Recent community efforts have sought to systematically evaluate the suitability of destructive \ac{SEM}-based inspection as a countermeasure against silicon Trojans.
One particularly valuable outcome is a public dataset~\cite{3.396Q7I_2022} associated with an IEEE S\&P 2023 publication~\cite{DBLP:conf/sp/PuschnerMBKMP23} that pairs backside \ac{SEM} images with anonymized \ac{GDS}~II layouts for multiple digital designs manufactured across a range of technology nodes (90\,nm, 65\,nm, 40\,nm and 28\,nm).
This dataset enables reproducible studies as well as benchmarking of algorithms aimed at detecting cell-level modifications.
It also helps to quantify how detection performance scales with shrinking feature sizes and increasing library complexity.
Early analysis of that dataset reported a clear degradation of detection performance in the most advanced node, as the majority of false positives and all observed false negatives were associated with identifications in the smallest technology generation~\cite{DBLP:conf/sp/PuschnerMBKMP23}.
The original authors suggested several plausible causes for this trend, including poorer image quality at smaller feature sizes as well as larger and more complex standard cell libraries in advanced nodes which increases the combinatorial pool of potentially similar-looking cells.

However, the evidence provided in the initial study did not allow a clean separation of these contributing factors.
In particular, it remained unclear to what extent the observed detection failures were due to the limits of \ac{SEM} imaging and stitching at nanometer scales or intrinsic properties of the cell libraries such as their overall complexity.
Furthermore, since only a few cell replacements have been introduced in each chip design, the original study may not have covered the continuous spectrum of detection difficulty (\eg, depending on how similar two exchanged cells look) sufficiently to draw generalizable conclusions.
Hence, it remains unclear how much the specific choices of cells that were replaced by Trojan surrogates have influenced the results.
Establishing a distinction between contributing factors is not merely academic.
If most detection errors are caused by imaging limitations, investments in higher-resolution equipment or alternative imaging techniques could materially improve detection.
Conversely, if indistinguishability is an inherent property of a subset of cells in certain libraries, then imaging quality improvements alone may not eliminate the risk.
Instead, security-aware library design, cell selection constraints, or architectural mitigation would be required.

\subsection{Our Contribution}
\label{sec:introduction:contribution}
This work provides a systematic and nuanced analysis of the S\&P 2023 dataset with the goal of quantifying the \enquote{Trojanizability} of standard cell libraries largely independent of image quality.
We define Trojanizability as the extent to which functionally different cells can be interchanged in fabricated chips without being visually distinguishable under conventional \ac{SEM}-based (backside) inspection.
We focus on functionally distinct cell pairs, as exchanging functionally equivalent cells does not introduce a meaningful attack surface.
In summary, we make the following contributions:

First, we develop an efficient, deterministic, and explainable similarity metric based on via placement.
Our metric leverages instance averaging across multiple cell occurrences to suppress imaging noise and artifacts.
This approach yields representative models of cell types that capture intrinsic visual characteristics rather than instance-specific distortions.
Using this metric, we perform comprehensive pairwise comparisons of all functionally different, same-width cell types across four technology nodes (\autoref{sec:robustmetric}).
This analysis directly addresses a central open question from prior work, namely whether detection failures stem from imaging limitations or from intrinsic properties of the underlying cell libraries.
Our results show that Trojanizability is largely an inherent property of the libraries themselves.
In particular, smaller process nodes exhibit a higher prevalence of visually similar but functionally distinct cells, with the smallest node showing markedly increased susceptibility.
Across all evaluated libraries, we identified cell pairs that are (close to) indistinguishable in backside \ac{SEM} imagery and are predominantly related through logical inversion (\eg, XOR vs. XNOR, BUF vs. INV, and TIEL vs. TIEH).
We refer to such pairs as \textit{invisible inversions}, which enable stealthy -- and in some cases effectively undetectable -- hardware Trojans.

Second, we evaluate how our metric can be used to enhance state-of-the-art Trojan detection (\autoref{sec:detection}).
Compared to the template- and via-mask-based methods of Puschner et al.~\cite{DBLP:conf/sp/PuschnerMBKMP23}, our approach achieves zero false negatives across all evaluated nodes and reduces false positives in three out of four cases.\footnote{The artifact can be found at \url{https://doi.org/10.5281/zenodo.19564962}}
It performs less effectively only for the 65\,nm images, where bright artifacts in the \ac{SEM} data are frequently misclassified as vias within individual cells.

Third, to demonstrate the implications of invisible inversions, we implement a stealthy privilege-escalation backdoor in an Ibex RISC-V core (\autoref{sec:casestudy}).
The resulting Trojan is functionally effective yet visually undetectable under \ac{SEM} inspection as performed in the provided dataset.

Our findings have two key implications.
Empirical evaluation of standard cell libraries for Trojanizability should become an integral part of the security review process for safety- or security-critical designs.
At the same time, because indistinguishable but functionally different cell pairs can be identified deterministically, they can be avoided during synthesis and place-and-route for high-assurance products, thereby achieving strong practical mitigation with minimal area or performance overhead. 
\section{Threat Model and Limitations}
\label{sec:threatmodel}

\subsection{Assumptions}
\label{sec:threatmodel:assumptions}
Consistent with prior work~\cite{DBLP:conf/sp/PuschnerMBKMP23} on the dataset underlying this study~\cite{3.396Q7I_2022}, we assume that the finished chip design leaving the \ac{IC} design house is benign and free of intentionally inserted Trojans or backdoors.
Consequently, we do not consider internal design stage threats such as untrusted employees within the design house, subverted \ac{EDA} tools, or malicious third-party \ac{IP} blocks introduced during development.
Our analysis therefore concentrates on external supply chain risks that arise after a design is finalized and ready to be submitted for fabrication.

\subsection{Adversary Scope}
\label{sec:threatmodel:scope}
We model adversaries who gain the opportunity to modify design data, masks, or other production inputs during the hand-off to, or processing by, an outsourced foundry.
A fully malicious foundry can, in principle, exercise almost arbitrary control over the silicon realization, ranging from subtle transistor-level doping changes to large-scale layout modifications, making the general problem of securing fabrication extremely challenging.
Prior literature has demonstrated that modifications at the process or doping level can alter circuit behavior without introducing obvious image anomalies~\cite{DBLP:conf/ches/SugawaraSFTHSF14}.

If the attacker model is essentially unrestricted, no practical defenses are feasible. To make the problem tractable and to focus on an attack class that is both realistic and compatible with standard manufacturing flows, we restrict the adversary to cell substitution attacks.
The adversary may replace one standard cell from the foundry's library or \ac{PDK} with another cell from the same library.
This constraint preserves the nominal physical and process characteristics expected by the mask generation and production pipeline and therefore does not require the foundry to deviate from normal mask creation or process steps in any obvious way.
Such substitutions can still be functionally potent while remaining subtle from a visual inspection perspective.

This model captures a broad and plausible set of attacker capabilities.
The adversary may be remote (\eg, an interceptor or attacker who tampers with the submitted \ac{GDS}~II files, or an adversary who compromises the foundry’s file storage or submission portal) or local (\eg, an individual foundry employee with access to intermediate files).
Importantly, cell substitution does not require collusion of the entire facility or conspicuous changes to manufacturing procedures.
A single individual with access to files can effect substitutions that are propagated undetected through conventional benign production pipelines.

\subsection{Comparison with Prior Work}
\label{sec:threatmodel:priorwork}

\begin{figure}[htb]
    \centering
    \begin{subfigure}[t]{.47\linewidth}
        \centering
        \includegraphics[height=3cm]{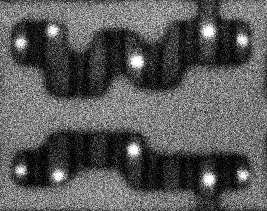}%
        \vspace{2pt}\\%
        \includegraphics[height=3cm]{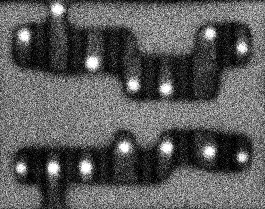}
        \caption{Trojan from~\cite{DBLP:conf/sp/PuschnerMBKMP23}.}
        \label{fig:oldvsnew_a}
    \end{subfigure}
    \begin{subfigure}[t]{.47\linewidth}
        \centering
        \includegraphics[height=3cm]{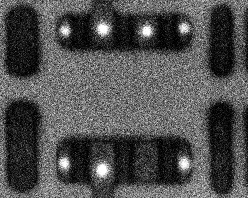}%
        \vspace{2pt}\\%
        \includegraphics[height=3cm]{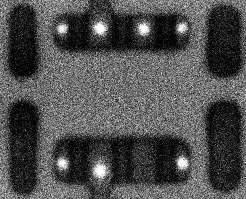}
        \caption{Possible stealthy Trojan.}
        \label{fig:oldvsnew_b}
    \end{subfigure}
    \caption{Comparison of a Trojan chosen by random cell replacement by~\citet{DBLP:conf/sp/PuschnerMBKMP23} (left) and a stealthy replacement that could have been performed (right).}
    \label{fig:oldvsnew}
\end{figure}

Our threat model is deliberately aligned with the practical insertions evaluated in the original study~\cite{DBLP:conf/sp/PuschnerMBKMP23}, but we consider a slightly more refined adversary.
Rather than replacing filler cells or arbitrary same-size cells, we focus on \textit{targeted substitutions of visually similar but functionally distinct standard cells} by knowledgeable adversaries.
\autoref{fig:oldvsnew} shows the difference in sophistication. The two cells depicted in~\autoref{fig:oldvsnew_a} may look similar at first glance, closer inspection reveals rather different patterns, most notably a different number of bright vias in each darker region. The two cells compared in~\autoref{fig:oldvsnew_b} cannot be distinguished by the same criterion. This refinement raises the bar for detection and more directly probes whether indistinguishability at the cell level can be exploited for stealthy Trojan insertion.

\subsection{Detection Capabilities}
\label{sec:threatmodel:detection}
We assume the same conceptual blue team capabilities as the original work.
Detection is performed by human analysts aided by \ac{SEM} imaging and automated image matching algorithms applied to the provided imagery.
Because we operate on the same published \ac{SEM} images and on the same anonymized layout data, our evaluation measures performance under equivalent imaging constraints and sampling strategies.

\subsection{Limitations}
\label{sec:threatmodel:limitations}

\textbf{Dependence on the published \ac{SEM} imagery:} Despite our best attempts at noise reduction, the analysis remains in part tied to the type (\eg, backside) and quality (\eg, contrast and spatial resolution) of the \ac{SEM} images in the public dataset.
When we report that two functionally different cells are indistinguishable, this statement is strictly with respect to the available imagery and the image processing primitives we employ.
It does not imply a fundamental impossibility of distinguishing the cells under all conceivable imaging conditions.
Yet, since we emphasize noise rejection in the process, not all imaging conditions limit us.

\textbf{Higher-resolution and process-sensitive imaging techniques:} Techniques exist that can reveal process- and doping-related differences beyond simple geometric layout (for example, methods that make dopant contrasts visible in electron microscopy~\cite{DBLP:conf/ches/SugawaraSFTHSF14}).
Such approaches can, in principle, disambiguate some of the cells that appear identical in our dataset.
However, obtaining uniformly high-quality dopant-sensitive \ac{SEM} imagery across an entire die is expected to be costly and operationally challenging, and therefore falls beyond the scope of the present study, which targets cost-effective, scalable inspection workflows.

\textbf{Process and mask deviations outside cell substitution:} We do not model adversaries that alter wafer processing parameters, mask alignment strategies, or mask-level features that are not representable as simple cell substitution from the same library.
While such attacks may be powerful, they frequently require noticeable deviations from standard foundry practice or more extensive collusion and instrumentation and thus fall outside the restricted adversary model we analyze.

\textbf{Electrical and functional side channels:} Our analysis is purely visual. We do not examine whether electrical testing, side-channel measurements (power, \acl{EM}), or post-silicon functional testing could detect the substitutions we identify.
In practice, such approaches may detect some classes of Trojans even when visual inspection fails.
While we assume this to be highly challenging with respect to the stealthy Trojans considered in this work, integrating such approaches with layout-level analysis is an important direction for future work.

Taken together, these assumptions and limitations outline a narrowly scoped but practically meaningful evaluation of Trojan insertion via visually similar standard cell substitution under realistic \ac{SEM}-based inspection techniques.
Our results indicate risks that are immediate for designers and auditors who rely on \ac{SEM}-based verification, while also acknowledging that stronger and more costly imaging as well as complementary testing approaches may potentially further reduce the attack surface identified here.
\section{Trojanizability Assessment}
\label{sec:robustmetric}

In this section, we systematically evaluate \emph{Trojanizability} across four technology nodes. 
We begin by revisiting the baseline detection approach applied to the dataset, before presenting our comprehensive methodology\footnote{The theoretical foundations are outlined in Appendix~\ref{app:theo}.}  for assessing cell-library Trojanizability.
Finally, we characterize the results across all four technology nodes, revealing clear patterns in how susceptibility to stealthy cell substitution varies with library and feature size.

\subsection{Baseline: Detection of Exchanged Cells}
Prior work on this dataset~\cite{DBLP:conf/sp/PuschnerMBKMP23} employed a scoring-based approach to detect 
cell instances that visually differ from their expected type, thereby identifying stealthy hardware Trojans based on exchanged standard cells. 
Cell-type representatives were selected in a rudimentary fashion (the first detected instance), and two scoring methods were employed: template matching followed by via-mask matching if needed.
Consequently, the metric relied on dataset-specific thresholds to determine whether a cell instance deviated from its expected type.

\subsection{A Robust Metric for Trojanizability} 
\label{sec:robustmetric:distinguishability}
To systematically determine whether two functionally different cells from the same library are difficult to distinguish -- and thus potential candidates for stealthy Trojan insertion -- the proposed method (a) constructs robust representatives for each cell type, resilient to imaging artifacts and misalignments, and (b) compares all functionally different cell representatives without relying on imagery-specific thresholds.

\subsubsection{Extracting Vias as Distinguishing Features}
\label{sec:robustmetric:distinguishability:extraction}
\begin{figure}[htb]
    \centering
    \begin{subfigure}{.24\textwidth}
        \centering
        \includegraphics[width=.9\linewidth]{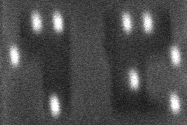}
        \caption{Cell from 90\,nm node}
        \vspace{0.5cm}
    \end{subfigure}%
    \begin{subfigure}{.24\textwidth}
        \centering
        \includegraphics[width=.9\linewidth]{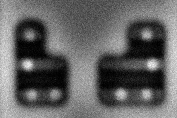}
        \caption{Cell from 65\,nm node}
        \vspace{0.5cm}
    \end{subfigure}
    \begin{subfigure}{.24\textwidth}
        \centering
        \includegraphics[width=.9\linewidth]{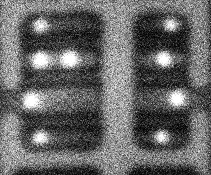}
        \caption{Cell from 40\,nm node}
    \end{subfigure}%
    \begin{subfigure}{.24\textwidth}
        \centering
        \includegraphics[width=.9\linewidth]{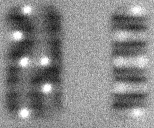}
        \caption{Cell from 28\,nm node}
    \end{subfigure}
    \caption{The subfigures show one exemplary cell instance from each technology node. Bright circular spots indicate vias, gray regions correspond to \acf{STI}, and the dark background represents the underlying silicon substrate.}
    \label{fig:all-cell-examples}
\end{figure}

The backside images of the poly layers across the four technology nodes reveal two consistent types of visible cell features -- vias and \acf{STI} -- as illustrated in \autoref{fig:all-cell-examples}.
Prior work on layout camouflaging suggests that functionally different cells rarely share identical via patterns~\cite{9039593, rajendran2013security}, although such cases cannot be fully excluded. 
Accordingly, we focus on vias as the primary distinguishing feature, since they can be reliably extracted even under noisy imaging conditions and efficiently represented as sets of two-dimensional points. 

To reliably detect vias in the 28\,nm tile images, which exhibit substantially higher noise levels and smaller brightness margins, we developed an enhanced detection method based on persistence analysis, a topological approach~\cite{DBLP:books/daglib/0025666, DBLP:journals/jossw/TralieSB18, TralieSaul2019LowerStarImageFiltrations}. 
Intuitively, this method can be viewed as \textit{flooding a landscape}: the grayscale image is treated as a topographical map where pixel brightness corresponds to elevation. 
The algorithm gradually drains the flooded landscape while tracking the emergence and merging of islands -- clusters of pixels -- recording their lifetimes. 
Pixels belonging to long-lived islands correspond to regions that remain bright relative to their surroundings. 
By retaining only these pixels, the method eliminates the need for fixed brightness thresholds and achieves robust via detection even under inconsistent illumination.

Because persistence analysis is over an order of magnitude slower than binary thresholding, we applied it selectively to the 28\,nm dataset, where it provided a clear improvement in detection reliability, while binary thresholding sufficed for the 90\,nm, 65\,nm, and 40\,nm images. 
For these technology nodes, we followed the via-detection procedure of \citet{DBLP:conf/sp/PuschnerMBKMP23}, which is based on simple grayscale thresholding followed by morphological erosion to remove small noise clusters and circular feature detection to locate via centers.

\subsubsection{Computing Cell-Type Representatives}
\label{sec:robustmetric:distinguishability:representatives}

To compare the similarity of different cell types, we construct a representative that characterizes all instances of a given type.
This representative must accurately capture the number and precise locations of all vias.

\paragraph{\textbf{Aligning the Vias of Cell Instances}}
Each cell type occurs in four possible orientations: untransformed, rotated by 180°, mirrored, or both. 
Following the preprocessing of Puschner~\etal, all instances are transformed into a canonical orientation to ensure comparability. 
Residual positional deviations may still occur due to slight inaccuracies in aligning the extraction bounding boxes with the image data, caused by stitching errors or minor stretching and rotation in the \ac{SEM} images.

\begin{figure}[htb]
    \centering
    \includegraphics[width=0.8\linewidth]{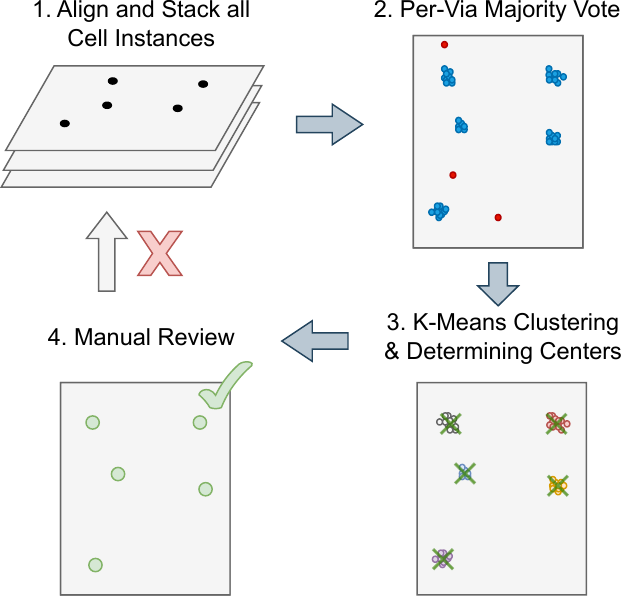}
    \caption{Process of computing a cell-type representative.}
    \label{fig:rep-computation}
\end{figure}

To compensate for these positional offsets and to ultimately generate robust representatives, 50 instances of each cell type are randomly sampled\footnote{Subsampling to 50 instances provides significant runtime improvement while maintaining sufficient data for robust representative generation.}, and -- in the first step (see \autoref{fig:rep-computation}) -- aligned.
We perform alignment pairwise among the 50 instances by trying for each pair all translations such that at least one via matches (with a via of the other instance).
We consider two vias matching if their distance is below \textit{half of a unit} of the chip's structure size (\ie, the minimum feature spacing within the technology node).
In this exhaustive search, we choose the translation with the maximum number of matching vias.
This provides a robust method to filter out any vias that exist only in a few instances, while also aligning all cells.

This method achieves robust results, as neither rotation nor scaling need to be considered once cells are normalized to canonical orientation. 
If one pair of actually corresponding vias aligns correctly, the remaining vias typically fall into place, and the approach remains tolerant to missing or noisy vias because partial matches still yield a meaningful alignment.

\paragraph{\textbf{Determining Via Amounts and Positions}}

In the second step, we apply a per-via majority vote across aligned cell instances to determine the precise number of vias for each cell type.
Points appearing in a majority of instances are retained as valid vias; isolated detections (likely imaging artifacts) are discarded. 
This filtering yields the definitive via count for each cell.
We then apply \textit{k-means} clustering to identify cluster centers, which serve as the via positions for the cell-type representative.

Finally, we perform manual verification of each representative through two validation steps. 
First, we visually compare the representative to actual cell instances to verify correct via positions and counts. 
Second, we align an independent subset of cell instances to the representative and confirm consistent alignment. 
If alignment issues or via inaccuracies are detected, we recompute the representative from a different randomly selected subset using progressively stricter majority-vote thresholds and re-validate.
Less than 10\% of representatives required correction, demonstrating that the initial approach is largely effective. 
This one-time manual verification step is crucial for ensuring accuracy of representatives, and is feasible given that cell libraries typically contain only hundreds of cell types -- representatives are then reusable across multiple analyses within the same cell library.

\subsubsection{Computing a Similarity Score}
\label{sec:robustmetric:distinguishability:score}

To quantify similarity between two cell-type representatives, we apply the alignment process described in \autoref{sec:robustmetric:distinguishability:representatives} and convert the result to a numerical score using the Jaccard distance adapted for via sets.
The similarity score ranges from \textit{1 (perfect similarity)} to \textit{0 (complete dissimilarity)} and is computed as the fraction of matched vias relative to the total number of vias in both cells. 
This metric penalizes misalignments: unmatched vias, missing vias, and vias displaced beyond half of a unit length all decrease the similarity.

\begin{figure}[htb]
    \centering
    \includegraphics[width=0.85\linewidth]{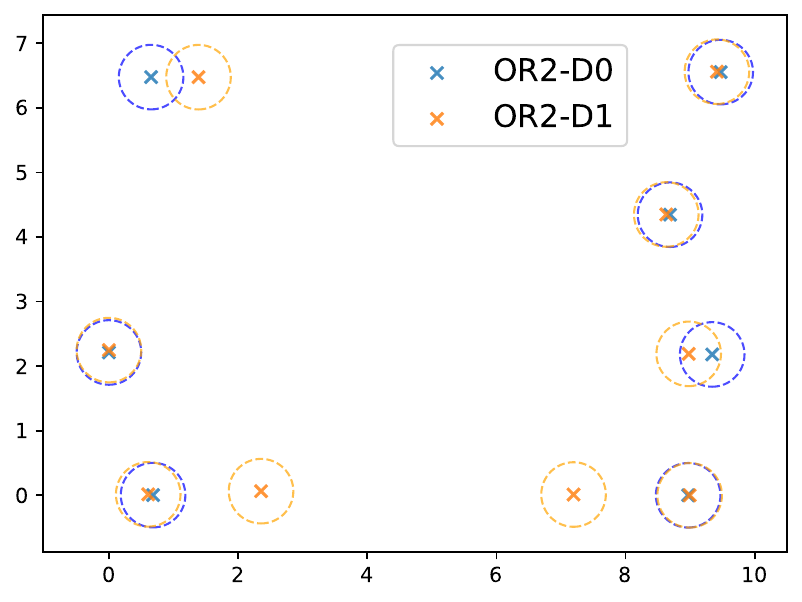}
    \caption{Similarity score calculation via Jaccard distance between \textbf{OR2-DO} and \textbf{OR2-D1}. Of 16 total vias, 12 align within the matching radius of half a unit length. Four unmatched vias contribute to a \textit{similarity score of 0.75}: two vias missing in OR2-DO (bottom) and two misaligned vias (top left, one from each cell).}
    \label{fig:jaccard}
\end{figure}

\subsubsection{Assessing Trojanizability via Similarity Scoring}
\label{sec:robustmetric:distinguishability:similarity}

To establish a notion of \emph{Trojanizability} for each technology node, we systematically determine similarity scores for all pairs of cell types, subject to two essential constraints: 
First, we restrict analysis to cell pairs of equal width, since cells of differing widths cannot be interchanged without violating layout constraints and design rules (multiple small vs. one large cell replacements are possible but expected to be easier to detect). 
Second, we assess only cell types that provide different functionalities, as exchanging cells with identical logic functions poses no functional security risk and thus no viable functional Trojan insertion point.
By computing these similarity scores across all valid cell pairs, we identify the most similar cell patterns within each technology node -- a critical foundation for our subsequent analyses of stealthy cell substitutions.
In Appendix~\ref{appendix:estimates}, we report estimates of computational costs and manual analyst effort for the key stages of our methodology.
\subsection{Trojanizability Across Technology Nodes}
\label{sec:robustmetric:results}

In the following, we present similarity scores for each technology node, followed by an analysis of the most similar cell pairs within each node.

\subsubsection{Similarity Scores per Technology Node}
\label{sec:robustmetric:results:scores}

\begin{figure*}[htb]
    \centering
    \includegraphics[width=\textwidth]{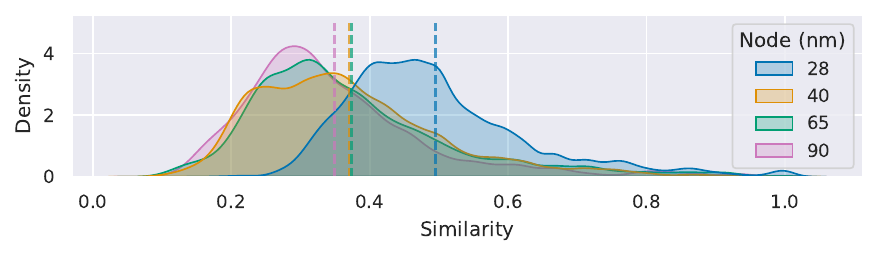}
    \caption{Kernel density estimates of similarity scores for all functionally distinct, same-width cell pairs across the four technology nodes. Higher scores denote higher visual similarity and thus greater susceptibility to stealthy cell substitution. Vertical lines mark mean values.}
    \label{fig:kde-plot-cell-scores}
\end{figure*}

\begin{figure}[htb]
    \centering
    \includegraphics[width=\linewidth]{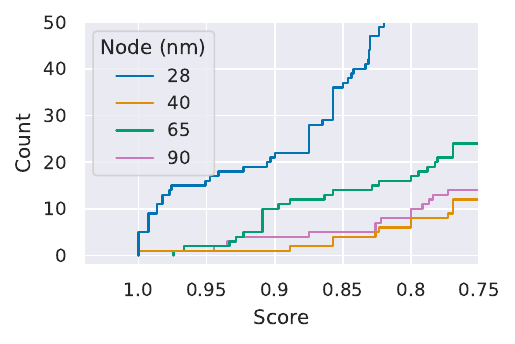}
    \caption{Cumulative count of valid cell pairs per technology node, ordered by decreasing similarity. This representation highlights how many highly similar cell pairs exist in each node, starting from the most similar.}
    \label{fig:ecdf-plot-of-scores}
\end{figure}

As shown in \autoref{fig:kde-plot-cell-scores}, the vast majority of cell types can be reliably distinguished, as most pairs of cell types exhibit low similarity.
The three larger technology nodes (40\,nm, 65\,nm, and 90\,nm) exhibit remarkably similar score distributions with nearly identical mean values of $0.37$, $0.373$, and $0.35$, respectively.
The 28\,nm node follows the same distributional shape but is substantially shifted toward lower scores, with a mean value of $0.5$.
This divergence becomes even more pronounced when examining the cumulative distributions in \autoref{fig:ecdf-plot-of-scores}:
the 90\,nm, 65\,nm, and 40\,nm nodes contain only very few cell pairs with highly similar or even identical via patterns, whereas the 28\,nm node exhibits substantially more such pairs.
Specifically, the 28\,nm node contains five pairs of functionally distinct cell types that are visually indistinguishable according to our metric -- a finding we examine in detail in the following section.

\subsubsection{(Most) Similar Cell Pairs per Technology Node}
\label{sec:robustmetric:results:similarpairs}
We now analyze the ten most similar cell pairs for each technology node (see Appendix~\ref{appendix:celltypes} for a complete overview), identifying the functional and structural patterns that characterize cell-level Trojanizability across our four different gate libraries.

\begin{figure}[hbt]
    \centering
    \begin{subfigure}[t]{0.32\linewidth}
        \centering
        \includegraphics[width=0.45\linewidth]{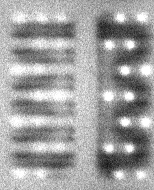}
        \includegraphics[width=0.45\linewidth]{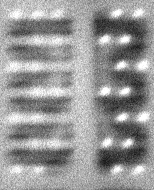}
        \caption{\centering \textbf{INV -- BUF}}
        \label{sub-fig:1st-most-similar-28nm}
    \end{subfigure}%
    \begin{subfigure}[t]{0.32\linewidth}
        \centering
        \includegraphics[width=0.45\linewidth]{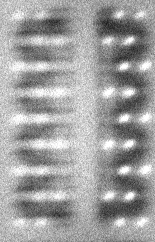}
        \includegraphics[width=0.45\linewidth]{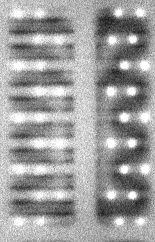}
        \caption{\centering \textbf{INV -- BUF}}
        \label{sub-fig:5th-most-similar-28nm}
    \end{subfigure}%
    \begin{subfigure}[t]{0.32\linewidth}
        \centering
        \includegraphics[width=0.45\linewidth]{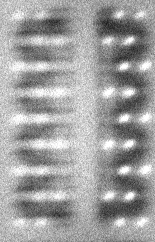}
        \includegraphics[width=0.45\linewidth]{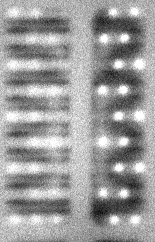}
        \caption{\centering \textbf{INV -- BUF}}
        \label{sub-fig:10th-most-similar-28nm}
    \end{subfigure}%
    \caption{Selected examples from the most similar cell pairs in the 28\,nm dataset: (a)~rank 1, (b)~rank 5, and (c)~rank 10, all showing INV--BUF variations.}
    \label{fig:top-10-similar-cell-pairs-28nm}
\end{figure}

The most similar cell pairs from the \textbf{28\,nm} cell library are predominantly buffer--inverter and XOR--XNOR variations.
\autoref{fig:top-10-similar-cell-pairs-28nm} illustrates three representative examples from the top ten most similar pairs.
The first and fifth most similar pairs (\autoref{sub-fig:1st-most-similar-28nm} and \autoref{sub-fig:5th-most-similar-28nm}) are indistinguishable according to our metric, with identical via patterns (similarity score of $1$).
The tenth most similar pair differs by only a single via (see bottom left of \autoref{sub-fig:10th-most-similar-28nm}; score $0.99$). 

\begin{figure}[htb]
    \centering
    \begin{subfigure}[t]{0.32\linewidth}
        \centering
        \includegraphics[width=0.45\linewidth]{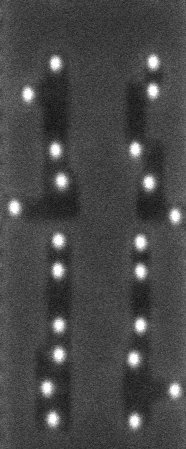}
        \includegraphics[width=0.45\linewidth]{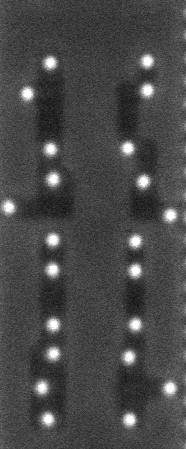}
        \caption{\centering \textbf{XOR -- XNOR}}
        \label{sub-fig:1st-most-similar-90nm}
    \end{subfigure}%
    \begin{subfigure}[t]{0.32\linewidth}
        \centering
        \includegraphics[width=0.45\linewidth]{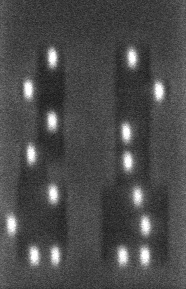}
        \includegraphics[width=0.45\linewidth]{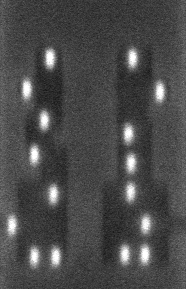}
        \caption{\centering \textbf{XNOR -- XOR}}
    \end{subfigure}%
    \begin{subfigure}[t]{0.32\linewidth}
        \centering
        \includegraphics[width=0.45\linewidth]{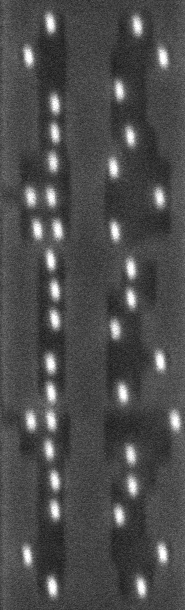}
        \includegraphics[width=0.45\linewidth]{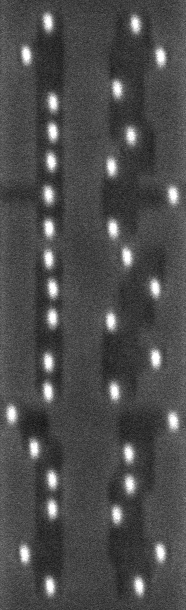}
        \caption{\centering \textbf{XNOR -- XOR}}
    \end{subfigure}
    \caption{Selected examples from the most similar cell pairs in the 90\,nm dataset: (a)~rank 1, (b)~rank 5, and (c)~rank 10, all showing XOR--XNOR variations.}
    \label{fig:top-10-similar-cell-pairs-90nm}
\end{figure}

The most similar cell pairs from the \textbf{90\,nm} cell library are predominantly XOR--XNOR variations, with one exception: EDF versus EDFQ (flip-flop with inverted output). 
Unlike the 28\,nm dataset, only a single pair achieves a similarity score of $1$ (see \autoref{sub-fig:1st-most-similar-90nm}). 
The remaining top-10 pairs are more distinguishable by our metric despite similar via patterns; the fifth and tenth most similar pairs score $0.87$ and $0.8$, respectively.

\begin{figure}[htb]
    \centering
    \begin{subfigure}[t]{0.32\linewidth}
        \centering
        \includegraphics[width=0.45\linewidth]{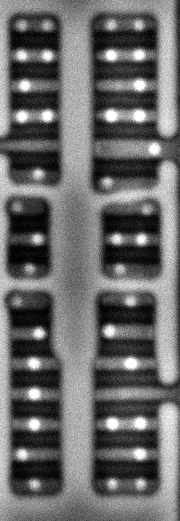}
        \includegraphics[width=0.45\linewidth]{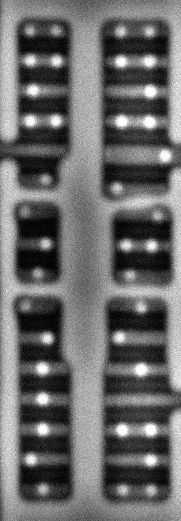}
        \caption{\centering \textbf{XOR -- XNOR}}
        \label{sub-fig:1st-most-similar-65nm}
    \end{subfigure}%
    \begin{subfigure}[t]{0.32\linewidth}
        \centering
        \includegraphics[width=0.45\linewidth]{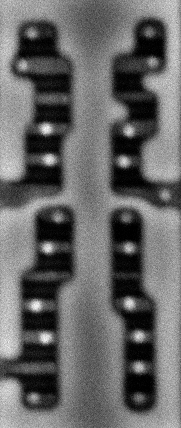}
        \includegraphics[width=0.45\linewidth]{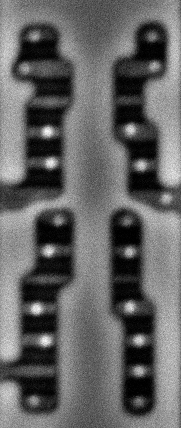}
        \caption{\centering \textbf{XOR -- XNOR}}
        \label{sub-fig:5st-most-similar-65nm}
    \end{subfigure}%
    \begin{subfigure}[t]{0.32\linewidth}
        \centering
        \includegraphics[width=0.45\linewidth]{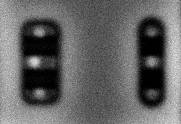}
        \includegraphics[width=0.45\linewidth]{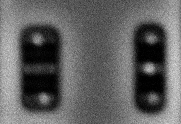}
        \caption{\centering \textbf{BUF -- NOR}}
        \label{sub-fig:10st-most-similar-65nm}
    \end{subfigure}%
    \caption{Selected examples from the most similar cell pairs in the 65\,nm dataset: (a)~rank 1, (b)~rank 5, and (c)~rank 10, showing XOR--XNOR and BUF--NOR variations}
    \label{fig:top-10-similar-cell-pairs-65nm}
\end{figure}

Several of the most similar cell pairs from the \textbf{65\,nm} cell library are XOR--XNOR variations, with scores of $0.97$ and $0.92$ for the first and fifth most similar pairs (see \autoref{fig:top-10-similar-cell-pairs-65nm}).
Beyond these, the top-10 pairs exhibit greater diversity, including BUF--NOR variations (tenth most similar pair, score $0.91$) and other uncommon pairings such as BUF--NAND and INV--Tie-Low (ties signal to ground). 
While no pair achieves a similarity score of $1$, the 65\,nm library exhibits the second-highest concentration of highly similar cells according to our metric (see \autoref{fig:ecdf-plot-of-scores}).

\begin{figure}[htb]
    \centering
    \begin{subfigure}[t]{0.32\linewidth}
        \centering
        \includegraphics[width=0.45\linewidth]{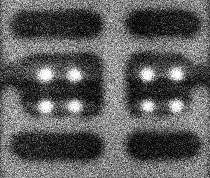}
        \includegraphics[width=0.45\linewidth]{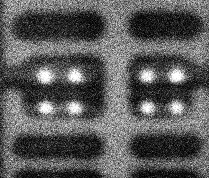}
        \caption{\centering \textbf{Tie-High -- Tie-Low}}
        \label{sub-fig:1st-most-similar-40nm}
    \end{subfigure}%
    \begin{subfigure}[t]{0.32\linewidth}
        \centering
        \includegraphics[width=0.45\linewidth]{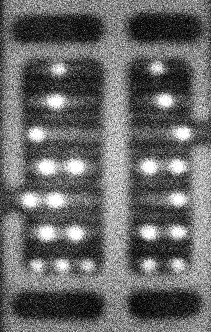}
        \includegraphics[width=0.45\linewidth]{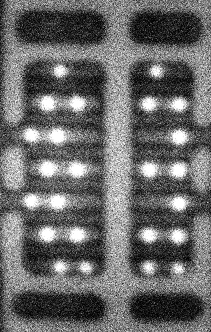}
        \caption{\centering \textbf{BUF -- INV}}
        \label{sub-fig:5st-most-similar-40nm}
    \end{subfigure}%
    \begin{subfigure}[t]{0.32\linewidth}
        \centering
        \includegraphics[width=0.45\linewidth]{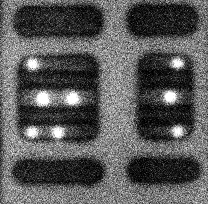}
        \includegraphics[width=0.45\linewidth]{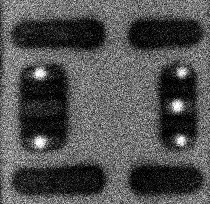}
        \caption{\centering \textbf{INV -- NOR}}
                \label{sub-fig:10st-most-similar-40nm}
    \end{subfigure}%
    \caption{Examples from the most similar cell pairs in the 40\,nm dataset: (a)~rank 1, (b)~rank 5, and (c)~rank 10.}
    \label{fig:top-10-similar-cell-pairs-40nm}
\end{figure}

The most striking result from the \textbf{40\,nm} dataset is a tie-high--tie-low pair with a similarity score of $1$ that is visually indistinguishable, as shown in \autoref{sub-fig:1st-most-similar-40nm}. 
In contrast, tie-high and tie-low cells in the other technology nodes typically differ in via count.
The 40\,nm library exhibits the fewest highly similar cells, yet simultaneously displays the greatest functional diversity among similar pairs: the fifth most similar pair (BUF--INV; score $0.83$) and tenth most similar pair (INV--NOR; score $0.77$) exemplify this breadth.
The top-10 most similar pairs further include AOI--IOA and OR--INOR variants. 
While all pairs except the most similar receive below-one scores and are theoretically detectable through thorough inspection, this functional diversity expands the attack surface available to adversaries.

\subsubsection{Summary of Findings}
\label{sec:detection:results:indistinguishable}

Across all four technology nodes, XOR--XNOR variations consistently appear among the top-10 most similar cell pairs, reflecting a broader pattern of functionally inverted or complementary cell types. 
The 40\,nm and 65\,nm libraries exhibit notably greater functional diversity in their most similar pairs compared to 28\,nm and 90\,nm.
Structurally, the 28\,nm, 40\,nm, and 90\,nm libraries each contain cell pairs with similarity scores of $1$, with 28\,nm being the most pronounced (five pairs). 
The 65\,nm library alone contains no such perfectly indistinguishable pairs under our metric. 
While the most similar 40\,nm pair is visually identical across \ac{SEM} images, other pairs with similarity score of $1$ in different nodes exhibit subtle structural differences that may contribute to visual distinguishability, though their practical utility for distinguishing instances from \ac{SEM} imagery remains unclear.

\section{Revisiting Trojan Detection}
\label{sec:detection}
Having identified cell pairs that are difficult to distinguish, we now analyze the likelihood of Trojans based on instance-level cell substitutions remaining undetected.
To this end, we turn towards individual cell instances and extend our similarity scoring approach from~\autoref{sec:robustmetric:distinguishability:score} with a classifier predicting whether individual cell instances are benign or Trojans. 
We then compare the detection performance of our refined method with the original detection method proposed by \citet{DBLP:conf/sp/PuschnerMBKMP23}, which utilizes either template matching, via-mask matching, or both.
With this experiment, we augment our initial notion of similarity with a practical analysis of detectability in which we focus particularly on the probability of false negatives, \ie, cell pairs that would likely remain undetected if exchanged.

\subsection{Refined Method for Trojan Detection} 
\label{sec:detection:method}

\subsubsection{Cleaning Up Cell Instances}
\label{sec:detection:method:cleanup}

Bounding boxes do not always align perfectly with cell boundaries on \ac{SEM} images.
To ensure complete cell extraction, we add a small safety margin to the bounding boxes~\cite{DBLP:conf/sp/PuschnerMBKMP23}.
Thus, extracted cell instances often contain vias belonging to neighboring cells. 
Template-based scoring methods are relatively robust to this contamination, as neighboring structure has minimal impact on the overall score.
Our via-position-based metric, by contrast, must explicitly filter neighboring vias; otherwise they strongly influence the score.

To address this issue when comparing via positions, each representative is assigned a box with dimensions based on the width and height of the respective cell type. 
When comparing a cell instance to its corresponding representative, it is first aligned, and then all of the vias outside of the representative's box are disregarded for the scoring process. 
If the cell instance is benign and most of the vias have been correctly detected, then this process will only remove vias stemming from neighboring cells, increasing its final score.
If the cell is a Trojan, then there are two possible scenarios:
First, the Trojan cell may align well with the representative, in which case the vias of the Trojan's neighboring cells are removed.
This minimizes the number of vias in the comparison and maximizes the impact of every non-matching via in the Trojan, thereby removing any chance of camouflaging a Trojan with vias of a neighboring cell.
This gives us the best chance to reproduce the theoretical difference between the two cells.
Secondly, if the Trojan does not align well with the representative, many of the vias will remain unmatched, resulting in a low similarity score.

\subsubsection{Scoring Approach}
\label{sec:detection:method:performance}
To derive a cell classifier from our new via-position based scoring method, we make a slight tweak to the original empirical Trojan definition of Puschner~\etal:
Under our threat model (\autoref{sec:threatmodel:scope}), we can define a cell substitution Trojan  as any cell that has a (significantly) \textit{better-fitting} representative than that claimed in the design specifications -- rather than merely as a cell that poorly fits its claimed representative.
This definition yields a classifier that is more robust against imaging noise in the instances.
To illustrate, an actual Trojan will score more similar to its true representative, and less similar to the claimed representative, triggering Trojan detection.
In contrast, a benign cell with a bogus via introduced by noise should be unlikely to have a corresponding via in any representative, thereby decreasing the similarity to all representatives in the same way and minimizing the risk of accidentally changing the best-fitting label.
A small caveat to this logic is that a cell instance may receive identical scores from multiple representatives.
In that case, it is only possible to guess the cell's true type.
This opens a trade-off between producing a large number of false positives, or flagging all such cells as benign and risking false negatives.
For the case that two types of cells are indistinguishable the evaluations would result in a 100\% false-negative rate, showing us that these two types of cells would have to be manually checked when searching for Trojans.

\subsubsection{Evaluation Protocol}
\label{sec:detection:method:scores}
Rather than relying on a case study of exchanged cells, to determine detectability, we perform this analysis exhaustively for all of the cell instances belonging to the types of the most similar cell pairs.
In our evaluation, we focus on the beta error since this is the critical concern when detecting Trojans: The chance of a Trojan slipping through any detection method should be as close to zero as possible.
The number of false positive detections is of secondary importance; it corresponds to the amount of manual verification required in the detection process and is better expressed in absolute numbers.
See Appendix~\ref{appendix:comparison} for the performance results of our new method on the Trojan detection challenge posed by Puschner~\etal

We implement our analysis as follows:
To test how detectable replacements of the most indistinguishable cell pairs are in practice, we compare all instances of each cell type with the representative of the other respective cell type in the pair.
We replicate the scenario in which a concrete instance of a cell has been replaced with a Trojan cell of the other type and is now being compared to the representative of the type it claims to be.  
By performing this analysis for all cell instances belonging to the two types, we can deduce how many cells were mislabeled as benign and how many were correctly labeled as a Trojan. This allows us to assign a false negative rate for each pair. 

\subsection{Results: Via Position Method vs. Baselines}
\label{sec:detection:results}

Comparing representatives as in~\autoref{sec:robustmetric:distinguishability:similarity} gives a good indication of how  similar two cell types are.
In contrast to the highly robust cell-type representatives, individual cell instances may be less accurate, as they are individually more susceptible to sample preparation and imaging artifacts.
Vias may not have been detected correctly, excess vias may have been falsely detected, or their positions are shifted.
Ideally, the vias would be detected at the exact same position for every cell, but since this is not the case, cells with vias in similar positions are more likely to go undetected when swapped.
We explore the detection rates that occur when swapping all instances of similar cell types to see how well they can be practically differentiated as in ~\autoref{sec:detection:method:scores}.
Concretely, we analyze the beta error for the most similar cell pairs, revealing how probable it is for a Trojan implemented by swapping these cell types to remain undetected.

\subsubsection{False Negative Rates of the Most Similar Cell Pairs per Cell Library}
\label{sec:detection:results:detection}

\begin{figure}[htb]
    \centering
    \includegraphics[width=\linewidth]{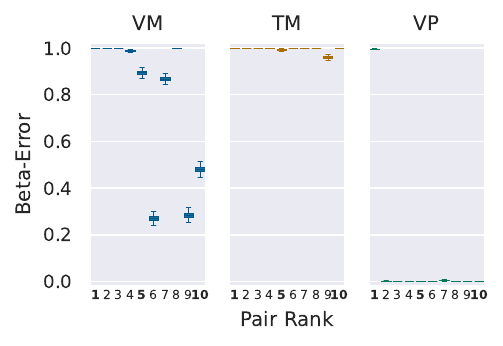}
    \caption{Beta errors of the ten most similar cell pairs in the 90\,nm gate library, comparing Via Matching~(VM) and Template Matching~(TM) from Puschner~\etal with our proposed Via Position~(VP) method.}
    \label{fig:beta-error-90nm}
\end{figure}

As shown in \autoref{fig:beta-error-90nm}, template matching fails to detect almost all substitutions of the ten most similar cell pairs in the \textbf{90\,nm} dataset.
Via mask matching shows improvement starting from the fifth most similar pair, while our method achieves near-zero false negatives from the second pair onward.
Only the most similar pair (XOR vs. XNOR), with an almost identical via pattern, remains practically indistinguishable, even with our metric.

\begin{figure}[htb]
    \centering
    \includegraphics[width=\linewidth]{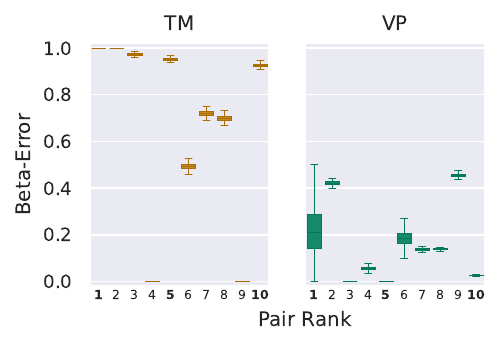}
    \caption{Beta errors in the 65\,nm gate library, comparing Template Matching~(TM) with our proposed Via Position~(VP) method.}
    \label{fig:beta-error-65nm}
\end{figure}

In the \textbf{65\,nm} dataset, detection performance is notably more challenging.
Our method achieves false negative rates of approximately $0.2$ or below for eight of the ten most similar cell pairs, but deteriorates to approximately $0.5$ for the second and ninth most similar pairs (see \autoref{fig:beta-error-65nm}).
Despite this mixed performance, our approach generally outperforms template matching, though template matching achieves better results for ranks 4 and 9 -- both exhibiting similar via patterns but marked differences in \ac{STI} structures.

The large confidence interval on the most similar pair is attributable to the extremely limited number of instances (3 and 2, respectively; see 
\autoref{tab:all:Top10Data}, 65\,nm section).
More fundamentally, the 65\,nm dataset exhibits image quality challenges: abundant false via detections due to bright imaging artifacts and high contrast between \ac{STI} and background, which benefits template matching methods.
Additionally, buffer cell types (part of ranks 7--10) exhibit high similarity to multiple other cell types, complicating via-based discrimination.

\begin{figure}[htb]
    \centering
    \includegraphics[width=\linewidth]{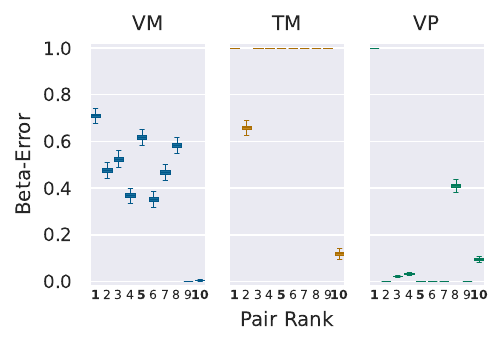}
    \caption{Beta errors in the 40\,nm gate library, comparing Via Matching~(VM) and Template Matching~(TM) with our proposed Via Position~(VP) method.}
    \label{fig:beta-error-40nm}
\end{figure}

In the \textbf{40\,nm} dataset, instances of the most similar cell pair remain indistinguishable across all evaluated methods (see \autoref{fig:beta-error-40nm}).
Template matching largely fails to detect substitutions of similar cell pairs, while via mask matching provides reliable detection only for ranks 9 and 10. 
Our method outperforms both baselines for all but the tenth most similar pair, where via mask matching achieves very low false negative rates.
However, our method shows degraded performance at rank 8 (and, to a lesser degree, rank 10), attributable to the abundance of functionally distinct cells with similar via patterns, as discussed in \autoref{sec:robustmetric:distinguishability:similarity}.

\begin{figure}[htb]
    \centering
    \includegraphics[width=\linewidth]{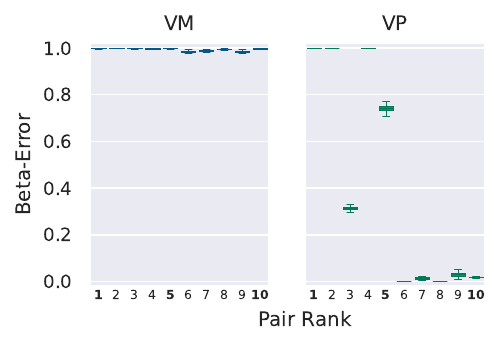}
    \caption{Beta errors in the 28\,nm node, comparing Via Matching~(VM) with our Via Position~(VP) method.}
    \label{fig:beta-error-28nm}
\end{figure}

In the \textbf{28\,nm} dataset, via mask matching fails to distinguish instances across all ten most similar cell pairs (see \autoref{fig:beta-error-28nm}).
As expected from the similarity scores of one between representatives of the five most similar pairs (see \autoref{sec:robustmetric:distinguishability:similarity}), our method also exhibits very high false negative rates for these pairs but achieves substantially improved reliability starting from the sixth pair.

\section{Case Study: Stealthy Hardware Trojans From Invisible Inversions}
\label{sec:casestudy}

To demonstrate the impact of our findings, we implemented and validated two silicon-level privilege-escalation Trojans in full \ac{ASIC} realizations of the Ibex RISC-V core\footnote{see \url{https://github.com/lowRISC/ibex}} in 65\,nm and 28\,nm technologies.
In detail, we have synthesized and placed and routed the Ibex architecture including \ac{PMP} to produce \ac{DRC} and \ac{LVS} clean \ac{GDS}~II files, targeting 50\,MHz working frequency under worst-case operating conditions (high temperature, low voltage), functionally verified through post-layout back-annotated timing simulation.
Since our designs are constructed from the same standard cell libraries as the ones of~\cite{DBLP:conf/sp/PuschnerMBKMP23}, we can pretend to have manufactured and imaged these chip designs without actually performing such steps, as \ac{SEM} pictures of the cells are included in the dataset~\cite{3.396Q7I_2022} with only few exceptions.

The Trojans were introduced by substituting a very small number of standard cells in the \ac{GDS}~II by their equivalents realizing the respective inverse function. In particular, we performed 3 XNOR--XOR substitutions in the 65\,nm variant and 2 INV--BUF substitutions in the 28\,nm variant. 
These changes cause the Boolean circuit that determines whether the current execution context has sufficient privilege to perform a requested operation to always evaluate to \emph{true}.
Importantly, our carefully chosen cell substitutions do not require any (re-)routing of signals, as the via patterns and connections on the lowest metal layers of original and substituted cells match exactly.
Layout-based inspection as well as many functional testing approaches will presumably struggle to detect this insertion.
Of course, once an application attempts to access privileged resources without permission which are actually reserved for firmware, bootloader or \ac{OS} kernels, the subversion would become evident.
An even stealthier Trojan design would require a dedicated trigger condition and need significantly more cell replacements and sophistication.

At a conceptual level, privilege checks in a processor core (\eg, RISC-V) are Boolean predicates combining mode bits, privilege status registers, and exception/interrupt state to decide whether a given instruction is allowed to access privileged resources, such as protected memory regions.
By altering the physical implementation of a small number of gates that feed the comparators responsible for this decision, we managed to cause the core to treat all execution as if it were in the highest privilege level.
A processor that treats all code as executing at the highest privilege level fundamentally breaks platform isolation.
Immediate practical consequences include i) arbitrary software (including untrusted user code or third-party applications) can access privileged \acp{CSR}, ii) secure boot and firmware integrity checks can be bypassed by altering the control flow that enforces them and iii) secrets stored in privileged memory regions (cryptographic keys, credentials, firmware images, or root secrets) can be read and exfiltrated by user-level code.

We have not performed such attacks nor manufactured the backdoored designs.
We emphasize that our writeup focuses on the evidentiary and conceptual aspects of the attack to explain what went wrong and why it is hard to detect with \ac{SEM} imagery of the published quality, rather than providing procedural instructions for reproducing or deploying such an attack in the wild.
Our aim is to inform designers, foundries, and auditors about realistic and subtle attack vectors so that practical mitigations can be implemented.

To demonstrate the feasibility of mitigating this risk, we evaluated the cost of excluding all problematic cell pairs from the 28\,nm design. 
Using standard \ac{EDA} tool \texttt{dont\_use} constraints on the five indistinguishable pairs (9 cell types in 8 variants each, resulting in 72 constraints), the overhead was negligible: $0.013\%$ area, $0.047\%$ timing, and $0.022\%$ power. 
In the other three technologies, fewer pairs are affected, causing presumably even smaller overheads.
This finding strongly suggests that practical defenses are available at minimal cost, a point we elaborate further in the subsequent discussion.

\section{Discussion}
\label{sec:discussion}

\subsection{Implications}
\label{sec:discussion:implications}

\subsubsection{Contemporary \texorpdfstring{\acp{PDK}}{PDKs} are Susceptible to Trojans}
\label{sec:discussion:implications:risk}

In all four cell libraries extracted from the public dataset, we identified pairs of cell types that are functionally different, but visually close to indistinguishable.
To an adversary choosing cell replacements wisely, this presents a lucrative toolbox for Trojan insertion in a highly surreptitious manner.
Exacerbating this issue, in all but the 65\,nm \acf{PDK}, we discovered cells with perfectly identical via placements.
To metrics focusing on via analysis, these cell pairs are perfectly indistinguishable in an information theoretical sense, \ie, detection of such cell replacements is impossible considering the type of imaging even in an ideal scenario assuming perfect sample preparation, zero noise, and no manufacturing variability.
Our detailed analysis of the template-matching-based method by Puschner~\etal shows that even computationally heavy methods taking into account features beyond vias will be largely unable to capture meaningful differences between those cells, granting an adversary a high likelihood of evading any type of optical inspection.
We consider it unlikely that a skilled human analyst fares better than the automated methods, especially considering the impracticality of manually checking each individual cell in large designs.

More complex sample preparation and imaging techniques are of limited help.
Front-side delayering to expose the lowermost metal layer may reveal visual differences in functionally diverse cells such as the INV--NOR pair encountered on the 40\,nm \ac{IC}.
However, for invisible inversions, such as the XOR--XNOR pair on the 90\,nm \ac{IC}, contacts are in identical locations and differences on this layer will be minimal or non-existent.
Additionally, the multitude of materials encountered in this approach renders sample preparation more difficult and costly.
Image analysis of this metal layer is concerned with polygons rather than point clouds, which would require defenders to extract vector representations from the cells -- an algorithmically more involved problem that remains the object of ongoing research~\cite{yu2022datadriven, rothaug2025advancing}.

A concerning trend in the dataset is the degradation of distinguishability with feature size, with the largest reduction observed moving from the 40\,nm to the 28\,nm design.
In the initial work on this dataset, this observation was partly attributed to poorer image quality, which certainly holds true:
For this image set, simple binary thresholding proved no longer sufficient for accurate via detection.
Noise and low contrast regions forced us to utilize persistence analysis to detect vias with sufficient precision.
However, not only did we need a more nuanced analysis for via detection, but despite the strong noise rejection that the abstraction from raw images to point clouds provides, average similarity scores were still considerably higher, indicating that there is less variation in the via patterns overall.
While the dataset only encompasses a single 28\,nm node, we hypothesize that the more regular structure is a consequence of the tighter integration and therefore applies in a more general sense.

Therefore, our answer to the open question posed by \citet{DBLP:conf/sp/PuschnerMBKMP23} is as follows:
Better sample preparation, imaging techniques, and post-processing will be beneficial in reducing noise and artifacts.
On their own, however, they do not represent a solution to the Trojanizability problem, as we found that cell designs themselves are less distinguishable with shrinking feature size.
We predict that this trend continues with further shrinkage, raising concerns about even higher Trojanizability in very advanced process nodes.
To avoid further increasing attackers' odds for success, interventions both on the process development and synthesis side are urgently needed.

\subsubsection{Trojan Verification is Reasonably Achievable}
\label{sec:discussion:implications:verification}

Our experiments have yielded empirical evidence that our improved metric is not only useful for classifying Trojanizability, but can also enhance Trojan detection efforts on real \ac{IC} designs.
Identical replacements aside, our metric performs well in detecting even minuscule differences between \ac{SEM} images of visually similar standard cells.
From the physical perspective, our analysis requires only backside imaging instead of expensive multi-step delayering.

When evaluated on the original detection experiment of~\citet{DBLP:conf/sp/PuschnerMBKMP23}, our metric detects all Trojans without false negatives, a result not achieved by the previously proposed techniques (see Appendix~\ref{appendix:comparison}).
However, our metric produces substantially more false positives in the 65\,nm node, where bright imaging artifacts are frequently misclassified as vias.
This discrepancy exposes a fundamental difference in failure modes: our via-position method is sensitive to \textit{spurious feature detections}, while template matching is sensitive to \textit{global appearance variation}, such as illumination gradients, \ac{STI} contrast, and stitching seams. 
Our method therefore excels when via extraction is reliable -- as in the 90\,nm, 40\,nm, and 28\,nm datasets -- because it abstracts away irrelevant pixel-level variation. 
Template matching captures complementary signals when via extraction is less reliable, for example in the 65\,nm images where high \ac{STI} contrast provides information our point-based representation may miss.
Together, the two methods exploit orthogonal information channels with complementary failure modes, supporting joint deployment.
Overall, our technique has runtime comparable to that reported by Puschner~\etal, with more time devoted to computing the initial cell representatives and less to the detection runs themselves (see Appendix~\ref{appendix:estimates}).
In all cases, the detection methods are efficient, deterministic, explainable and require little human intervention, while relying on well-founded statistical principles and established algorithms.

While our current sample size of four evaluated standard cell libraries is obviously limited, we are optimistic about the prospects of successfully applying the same technique to other, especially more advanced, feature sizes and libraries.
Our method should be applicable as long as vias remain a discriminating identifier of the underlying cell type and remain clearly distinguishable, regardless of shape or size, when averaging over sufficiently many instances.

\subsection{Recommendations}
\label{sec:discussion:recommendations}

Our findings imply the Trojanizability of all four tested standard cell libraries, some with fully indistinguishable replacements.
Using those libraries as-is in high-assurance \ac{IC} designs puts the designers and the end users of the devices built around such chips at a serious risk of highly surreptitious hardware Trojans.
In the following, we suggest immediately actionable mitigation steps \ac{IC} designers might take, and prompt \ac{PDK} designers to explicitly harden future cell libraries against stealthy cell replacement attacks.

\subsubsection{Empirical Validation of \texorpdfstring{\ac{PDK}}{PDK} Trojanizability is Prudent}
\label{sec:discussion:recommendations:validation}

Our metrics for via-based cell similarity analysis provide a robust toolbox to benchmark \acp{PDK} for distinguishability.
Designers of high-assurance \acp{IC} seeking to select a manufacturing process should favor those standard cell libraries that have fewer highly similar cells, and particularly avoid those libraries that include fully indistinguishable cells.
Doing so increases the chances of success for Trojan detection, and builds confidence in proofs of Trojan freeness, as we observed that chances of false-negative detections decrease with higher dissimilarity.

Should a situation arise where \ac{IC} designers cannot make that choice, a tradeoff between circuit optimization and distinguishability may be possible.
In this case, designers may benchmark a \ac{PDK} for similarity, identify cell pairs that are both functionally different and highly similar, and exclude both cells in the cell pair from use in the circuit design.
For instance, in the 28\,nm \ac{PDK}, it would be prudent to exclude the cell types making up pairs 1--5, resulting in a total of nine excluded cell types.
Those pairs yielded extremely low similarity scores and our detection method produced a significant number of false negatives.
We deem this exclusion approach viable as there generally exist multiple possible circuit realizations of combinational logic functions, and potentially difficult to exclude sequential standard cells such as flip-flops posed low risk for Trojan insertion in our results.
As the excluded cells remain part of the manufacturer's \ac{PDK}, an adversary retains the capability to use these cells in replacements.
However, as there would not exist any sufficiently similar cells in the design as viable stealthy targets, such replacements would become identifiable.
When problematic cells can be effectively barred from use in this manner, Trojan verification is feasible and efficient.

We suggest that based on our method there exist two viable approaches to \ac{PDK} benchmarking:
One option requires following the sample preparation and imaging procedure outlined by Puschner~\etal to obtain cell images from a device using the \ac{PDK} under study, after which via positions can be recovered using our via extraction mechanism.
Should designers have access to detailed \ac{PDK} information including via placement within standard cells, this could enable a more cost-effective alternative:
The similarity scoring mechanism we propose is invariant towards the source of the cell representatives used, such that representatives could be directly derived from the \ac{PDK}, avoiding the need for reverse engineering a physical sample.
In either case, analysis of a standard cell library has to be performed only once to determine which cells need to be avoided.

\subsubsection{Calling for Trojan-resistant \texorpdfstring{\acp{PDK}}{PDKs}}
\label{sec:discussion:recommendations:design}

Beyond \ac{IC} designers benchmarking existing \acp{PDK}, we call on manufacturers to consider cell-level Trojans in the development of future standard cell libraries.
In a hardened \ac{PDK}, a minimum visual difference between cell types should be ensured as a design rule.
As a strict minimal requirement, no two cell types may share the same via pattern.
More formally, the cell's via patterns in such a library could be regarded as a form of inherent watermarking rendering the library friendly towards backside imaging.
Similar work has proposed the use of nanoplasmonics to make cell types clearly distinguishable with optical microscopy under near-infrared illumination~\cite{zaraee2020gatelevel}, or to add other optical features to filler cells to improve image contrast~\cite{zhou2021hardware}.

Expanding this threshold to include non-identical but highly similar cells will enable more cost-effective verification with less precise imaging, at the cost of further restricting the \ac{PDK} design space.
Effectively, such a cell library would eliminate the threat of nearly invisible cell replacements, instead forcing adversaries to revert to clearly detectable modifications.
A carefully hardened \acp{PDK} would enable full verification of cell-level Trojan-freeness based on the proposed fast and robust detection method combined with the imaging techniques described by Puschner~\etal

\subsubsection{Procedural Controls and Practical Verification}
\label{sec:discussion:recommendations:additional}

Given finite verification budgets, we recommend the following prioritization for high-assurance designs. 
\textit{Procedural controls} -- including authenticated and integrity-protected \ac{GDS} submission, strict access controls, and traceable change logs -- should form the baseline for all technical measures, as they reduce opportunities for tampering before any inspection becomes necessary.
\textit{Library Trojanizability assessment} combined with synthesis constraints is a one-time, low-cost design step that eliminates the attack surface for invisible inversions at negligible overhead and should therefore be standard practice.
\textit{Functional and electrical testing}, including power side-channel analysis, is non-invasive and inexpensive relative to physical inspection and should be applied routinely. 
However, it provides only limited assurance against the threat class considered here: a Trojan with a sufficiently rare trigger condition can be engineered to remain functionally undetectable under realistic test coverage.
As a result, such testing should be treated as a complement to, rather than a substitute for, physical verification.
\textit{\ac{SEM}-based cell-level inspection} using our via-position metric, complemented by template matching where via extraction is unreliable, is substantially more expensive but directly targets cell-substitution attacks regardless of trigger conditions. 
Where the required assurance justifies the additional cost, more advanced imaging techniques such as dopant-sensitive \ac{SEM} can provide an additional layer of confidence.

\subsection{Related Work}
\label{sec:discussion:relatedwork}
Both conceptually and methodologically, individual aspects of this work have been treated in prior research.
In the following, we review works that have contributed to Trojan detection both on the level of standard cells and full layer imagery, as well as approaches to via detection.

\citet{liu2024novel} recently developed a Trojan detection methodology comparing \ac{SEM} images against original design files.
Focusing on vias, they derive cell representatives from \ac{GDS}~II files, requiring detailed access to the standard cell layout information of the \ac{PDK}.
While this work's main focus is on Trojan detection, the authors also present a cursory cell similarity analysis on a 55\,nm node.
Notably, while they did not observe indistinguishable via patterns in their study, the authors do caution that many cells bear optical similarity.
\citet{wilson2019novel} made similar remarks on a 32\,nm and 90\,nm cell library which they recovered from \ac{SEM} images on the via level.
In the 90\,nm node, they further remarked compatibility concerns with their strictly rasterized feature vectors, as the cell library was not following a fully rasterized pattern.
We believe that our more flexible metric poses a viable solution to this challenge.

A different stream of work focused on detecting Trojans by matching the active areas of transistors on \ac{SEM} images.
Various authors based their analyses on direct chip-to-chip or chip-to-\ac{GDS} comparison~\cite{courbon2015semba,courbon2015high,vashistha2018detecting}, forgoing any individual treatment of standard cells.
Later work made use of template matching to retrieve standard cells from \ac{SEM} images, enabling the direct analysis of cell replacements~\cite{courbon2020practical}.
In more recent years, supervised \ac{ML} approaches were introduced for cell extraction both on the active layer~\cite{lin2023sem2gds} and other layers accessible through different sample preparation techniques~\cite{bao2014application}.
\citet{vashistha2022detecting} improved training of \ac{ML} models by embedding specially protected training cells into the design, allowing training of detection models in situ without relying on golden samples.
On the level of individual visual features, such deep learning frameworks have also found use in object detection for extracting via positions, among other features~\cite{cheng2023domainintegrated,cheng2018hybrid,yu2022datadriven,lin2020deep}.

Lastly, some prior research has investigated concrete examples for meaningful Trojans in RISC-V microprocessors.
\citet{dharsee2023jinn} have presented a case study for logic insertion in an earlier design stage with the goal of circumventing software-based security controls.
On the level of cell-level manipulations, \citet{parvin2023trojand2} have shown the viability of inserting low-overhead Trojans into empty space on the \ac{IC} to modify program control flow.

\section{Conclusion}
\label{sec:conclusion}

Puschner~\etal's 2023 dataset provides one of the few open resources enabling reproducible study of hardware Trojan detectability from \ac{SEM} imagery. 
Building on this foundation, we shift the focus from identifying specific Trojans to understanding the inherent susceptibility of standard-cell libraries to visually undetectable cell substitutions -- a property we term as \emph{Trojanizability} of a cell library.

Addressing this problem required a new methodology. 
We introduced a new metric designed to quantify how easily functionally distinct cells can be confused in \ac{SEM} images. 
Our approach leverages extensive instance averaging to minimize the impact of imaging noise, thereby isolating the inherent distinguishability properties of the cell appearance. 
Even under these idealized conditions, susceptibility to substitution grows significantly with technology scaling, with the 28\,nm node standing out as the most challenging due to the higher density of visually similar cells in its library.

While not our primary goal, our metric also surpasses prior approaches by Puschner~\etal in straightforward Trojan detection while maintaining comparable computational cost.
Concerningly, our analysis reveals that in every evaluated technology, cell pairs exist that implement different logic functions yet are effectively indistinguishable in the \ac{SEM} images available in the dataset.
Nearly all such pairs -- and most high-similarity cells more generally -- are logical inversions of one another.
These \emph{invisible inversions} create a powerful attack vector: an adversary can substitute a cell with its inverted counterpart to alter functionality without modifying geometry or routing.

To demonstrate the real-world consequences of this vulnerability, we realized a privilege-escalation Trojan in two full \ac{ASIC} implementations of an Ibex RISC-V core using only indistinguishable cell substitutions.
The resulting modification bypasses privilege checks entirely while remaining visually undetectable under \ac{SEM} imaging of the fidelity of the S\&P 2023 dataset.

Our results show that Trojanizability is both real and severe.
Visually indistinguishable but functionally distinct cells exist in all evaluated libraries, and their presence enables a class of hardware Trojans that are exceptionally difficult to detect using current \ac{SEM}-based inspection flows.
Fortunately, mitigation is straightforward: Once the Trojanizability of a cell library is assessed, protecting security-critical designs becomes simple.
Library users can avoid cells involved in indistinguishable pairs, incurring negligible overhead even in the most vulnerable technology node examined. 
We therefore recommend that designers of high-assurance \acp{IC} incorporate library-level Trojanizability evaluation into standard design and verification workflows, and suggest that future library development consider visual distinguishability as a design requirement.
To facilitate adoption of these recommendations, we open-source our implementation of the via extraction and similarity analysis pipeline as an accompanying artifact.

\section*{Acknowledgments}
Thorben Moos is a postdoctoral researcher of the Belgian Fund for Scientific Research (F.R.S.-FNRS).
This work was supported by the Deutsche Forschungsgemeinschaft (DFG, German Research Foundation) under Germany’s Excellence Strategy -- EXC 2092 CASA -- 390781972, and by the \href{https://rc-trust.ai}{Research Center Trustworthy Data Science and Security}, one of the Research Alliance Centers within the \href{https://uaruhr.de}{UA Ruhr}.

\bibliographystyle{IEEEtranN} 
\bibliography{IEEEabrv,bibliography}

\section*{Ethics Considerations}
This work analyzes the visual indistinguishability of standard cells within their libraries and explores the associated security implications. We have taken the following steps to communicate our findings responsibly:

1. \textbf{Identification, Communication, and Mitigation of Visually Indistinguishable Cells.}
Our analysis reveals that certain functionally distinct cells in contemporary standard cell libraries are visually indistinguishable under realistic backside SEM inspection, creating a potential avenue for stealthy hardware Trojans. 
We consider it important to highlight this class of risks so that defenders can account for them, while ensuring that no specific product or vendor is placed at additional risk.
To support defenders, we outline simple and practical mitigations such as excluding problematic cell types in high-assurance designs and encouraging PDK development practices that improve visual distinguishability. 
We also release our analysis tooling as open-source software to support defensive analysis, enable independent verification, and improve detection workflows.
Since the underlying dataset anonymizes all libraries and reveals no commercial provenance, vendor notification is neither feasible nor applicable.

2. \textbf{Construction of Stealthy Hardware Trojans.}
In our case study, we implemented proof-of-concept Trojans solely within controlled research environments. We intentionally refrain from releasing low-level implementation details, scripts, targeted signals, or exact modification procedures that would enable direct replication. All descriptions are conceptual and intended only to illustrate the feasibility and impact of indistinguishable cell substitutions. No fabricated silicon was produced for Trojanized designs.

\section*{LLM Usage Considerations}
LLMs were used for editorial purposes in this manuscript, and all outputs were inspected by the authors to ensure accuracy and originality.

\appendices

\section{Theoretical Foundations of Via-Based Scoring}
\label{app:theo}

This appendix details the mathematical and statistical foundations of the feature extraction and classification techniques used in Sections~\ref{sec:robustmetric} and~\ref{sec:detection}. 
In doing so, we build toward showing that, given via-layer information and a known candidate substitution, the separation margin is an optimal test statistic.

\subsection{Feature Selection and Translation Invariance}

The primary goal of our methodology is to provide theoretical guarantees on the noise thresholds the system can handle while minimizing manual verification effort. 

Previous approaches in the literature established translation invariance using the correlation coefficient by shifting full-cell images to identify the maximum alignment. While functional, dense full-image calculations are computationally expensive, difficult to analyze theoretically, and highly sensitive to non-structural variations. We address this by reducing images to point sets that represent spatial information via coordinates, thereby isolating the cell's semantic layout. Alignment then becomes a rapid point-set matching problem rather than a dense matrix operation.

To robustly extract these coordinates (\autoref{sec:robustmetric:distinguishability:extraction}), we use \textit{persistent homology}, a topological method that tracks how long features persist across brightness thresholds. When the image is represented as a \textit{cubical complex}, true vias are extracted as features with high persistence ($\pi := b - d$). These vias appear as local brightness maxima because the dense materials scatter electrons. They are born at high brightness thresholds ($b$) and merge with the background only at much lower thresholds ($d$). 

\textbf{Noise and Stability:} Intrinsic electron microscopy noise creates high-frequency fluctuations that yield short-lived topological features. Conversely, true vias produce stable signals despite large variations in lifetime. Extraction stability is backed by the \textit{Cohen-Steiner Theorem}, which guarantees that small image intensity changes lead to only small changes in extracted topological features.

\subsection{Spatial Noise and the Distance Measure}

Once extracted, features are compared using a distance metric. We model two primary sources of spatial noise:

1. \textbf{Structural Artifacts (The Matérn Cluster Process).} Features like Shallow Trench Isolation corners can survive extraction. Arising from physical chip structures rather than sensor noise, they tend to cluster. We model them using a \textit{Matérn Cluster Process}, a statistical model for random points that appear in dense clumps around unobserved parent locations. Because they rely on flat-gradient regions, their extracted positions are more varied than those of true vias.

2. \textbf{Spatial Jitter (The Rayleigh Distribution).} In the intensity terrain, vias appear as steep peaks. Stochastic noise causes small, localized shifts in the coordinates of their local maxima. As we expect an exponential decay of the probability density over noise brightness, we model the 2D spatial displacement as a Gaussian random variable, and the shift magnitude follows a \textit{Rayleigh distribution}.

Because the empirical Rayleigh variance is strictly smaller than the minimum distance between design grid points, large shifts are highly unlikely. Thus, we set the tolerance radius to half a unit length and utilize the \textit{Undirected Jaccard Metric} bounded by this radius. Any detection beyond this is penalized, making the Jaccard distance highly sensitive to layout alterations (\autoref{sec:robustmetric:distinguishability:score}).

\subsection{Representative Construction \& Error Scaling}

To classify instances, we construct a clean ground-truth template, or \textit{representative}, for each cell type (\autoref{sec:robustmetric:distinguishability:representatives}). Translation invariance is established by pairwise aligning point clouds to minimize the distance between them. Repeated for multiple instances, these alignments form clusters used for significance analysis.
Because large spatial shifts are restricted by Rayleigh bounds, unmatched points likely originate from randomly detected peaks (Matérn artifacts) rather than shifted true vias. We can therefore model the probability of a cluster corresponding to a true via independent of other detections outside the threshold radius.

To filter artifacts, we overlay a subsample of instances (e.g., $k=50$). Given the probability $p$ of detecting a true via at a specific grid location, the total detections across $k$ samples follow a \textit{Binomial distribution}. Because true via detection probability significantly exceeds $50\%$, a majority vote reliably confirms them. Conversely, Matérn artifacts have a detection probability well below $50\%$. Consequently, the misidentification probability decays exponentially with $k$, making representative errors vanishingly small at $k=50$.

\subsection{Classification, Bias, \& Verification Confidence}

Given an instance $V_\mathrm{obs}$ and a representative $R$, as we can now assume $R$ to be error-free and the observation of vias to be independent and identically distributed, we model the likelihood as:
$$P(V_\mathrm{obs}|R) \propto (p_\mathrm{match})^{|V_\mathrm{obs} \cap R|} (p_\mathrm{miss})^{|V_\mathrm{obs} \cup R| - |V_\mathrm{obs} \cap R|}.$$

Classification is then formulated as a \textit{Likelihood Ratio Test}, in which we compare the probability of the observed data under the provided label $R_L$ with that under the alternative $R_\mathrm{alt}$.
The log-likelihood ratio simplifies algebraically to a form proportional to the difference in undirected Jaccard distances, defined as the \textit{Separation Margin} $\Delta$ (\autoref{sec:detection:method:scores}). Crucially, when evaluating a specific known substitution (e.g., XOR vs. XNOR), this reduces to a simple hypothesis test, and the \textit{Neyman-Pearson lemma} applies; thresholding on the separation margin is thus the optimal decision rule. For unknown substitution types, the best-fitting alternative $R_\mathrm{alt}$ is found via \textit{Maximum Likelihood Estimation} over the cell library, yielding a \textit{Generalized Likelihood Ratio Test}.

\textbf{Prior Bias for False Positive Reduction:} In our setting, the provided label $R_L$ is untrusted but usually correct. To prevent noise from triggering false anomalies and to mitigate the multiple comparisons problem, we apply a prior bias $\eta$ to $R_L$. The system suggests an alternative label $R_\mathrm{alt}$ only if its Jaccard distance is substantially better. One could start with a target false-negative rate and compute a threshold that minimizes manual effort, given the probabilities; however, this is infeasible in practice because precise estimation of the tails of the underlying probability distributions is required. \emph{Sorting the dataset by $\Delta$ and inspecting the lowest-scoring ``long tail'' yields the best achievable false-negative rate for a given manual effort budget.}

\textbf{Confidence Parameters and Practical Learnings:}
The confidence in the method is governed by three primary parameters that dictate when human intervention is necessary:
\begin{itemize}
    \item \textbf{Geometric uniqueness and via count ($|V|$)}: The statistical reliability of the Jaccard distance scales with the number of vias ($|V|$) in a cell given spatial independence outside the tolerance radius. Cells with high via counts exhibit high geometric uniqueness, consistent with our statistical guarantees. Conversely, sparse cells lack sufficient spatial information to distinguish true matches from random artifact alignments and therefore require manual scrutiny.

    \item \textbf{Number of cell instances ($|I|$)}: As demonstrated by the binomial error scaling, separating true vias from artifacts requires a sufficient number of instances per cell type. For rare cells lacking sufficient instances to cleanly filter out noise during the majority-vote construction, the generated ground-truth representative may be flawed and must be reviewed manually. Fortunately, reviewing flawed representatives for rare cells incurs minimal manual effort. 

    \item \textbf{The Separation Margin ($\Delta$)}: A large, positive $\Delta$ indicates that the assigned label fits the empirical point cloud significantly better than any alternative. We constrain the alternative-label search space to cell types with similar physical dimensions to reduce unnecessary computation. Dissimilar cells will naturally score a high $\Delta$ and thus never trigger a manual review, making this structural constraint a safe optimization for computational efficiency.
\end{itemize}

\newpage
\onecolumn

\section{Detailed Empirical Results}

\subsection{Estimates of Computational Cost and Manual Analyst Effort}
\label{appendix:estimates}
\leavevmode 
\begin{table}[H]
    \centering
    \caption{Estimates of the computational cost and manual effort required for the detection process. All computations were run on a laptop using a \textit{12th Gen Intel(R) Core(TM) i7-12700H} CPU with 16GB of RAM.}
    \begin{tabular}{c|ccccc}
       \toprule
       Node  &  Extracting Vias & Computing Representatives & Comparing Representatives & Manual Review & Trojan Detection\\
       \midrule
       90 nm & 1 h & 1 h & $<$ 0.25 h & 2 h -- 3 h & 0.5 h  \\
       65 nm & 1.5 h & 1 h & $<$ 0.25 h & 2 h -- 3 h & 1.5 h\\
       40 nm & 2.5 h & 1.5 h & $<$ 0.25 h & 2 h -- 4 h & 1 h\\
       28 nm & 20 h & 2 h & $<$ 0.5 h & 2 h -- 4 h & 1 h\\
       \bottomrule
    \end{tabular}

    \label{tab:estimates}
\end{table}
\smallskip

\subsection{Supplementary Data for the Top Ten Most Similar Cell Types per Technology Node}
\label{appendix:celltypes}
\leavevmode 

\begin{table}[H]
    \centering
    \caption{Logic functionality, number of instances in the underlying dataset, and similarity scores of the ten most similar, functionally distinct, same-width cell pairs across the four technology nodes under study.}
    \label{tab:all:Top10Data}
    \footnotesize 
    \begin{tabular}{cl|cccccccccc}
        \toprule
        Node & Pair Rank & 1 & 2 & 3 & 4 & 5 & 6 & 7 & 8 & 9 & 10 \\
        \midrule
        \multirow{5}{*}{\textbf{90\,nm}} 
        & Cell Function (1) & XOR & XNR & XNR & XNR & XNR & XNR & XOR & EDF & XOR & XNR \\
        & Cell Function (2) & XNR & XOR & XOR & XOR & XOR & XOR & XNR & EDFQ & XNR & XOR \\
        & \# Instances (1) & 3106 & 808 & 12 & 339 & 507 & 5192 & 3106 & 5 & 2778 & 12 \\
        & \# Instances (2) & 5192 & 140 & 18 & 6 & 2778 & 394 & 101 & 7 & 339 & 140 \\
        & Similarity Score & 1.00 & 0.94 & 0.94 & 0.93 & 0.88 & 0.83 & 0.83 & 0.82 & 0.80 & 0.80 \\
        \midrule
        \multirow{5}{*}{\textbf{65\,nm}} 
        & Cell Function (1) & XOR & XOR & ND & XNR & XOR & INV & BUFF & BUFF & BUFF & BUFF \\
        & Cell Function (2) & XNR & XNR & I-NAND & XOR & XNR & Tie-Low & NAND & NAND & NOR & NOR \\
        & \# Instances (1) & 3 & 818 & 177 & 723 & 4693 & 131 & 13235 & 13235 & 13235 & 13235 \\
        & \# Instances (2) & 2 & 2898 & 6 & 14 & 3077 & 2 & 239 & 13607 & 18 & 4890 \\
        & Similarity Score & 0.97 & 0.97 & 0.93 & 0.93 & 0.92 & 0.91 & 0.91 & 0.91 & 0.91 & 0.91 \\
        \midrule
        \multirow{5}{*}{\textbf{40\,nm}} 
        & Cell Function (1) & Tie-High & XOR & I-NOR & OAI & BUFF & XNR & OR & IOA & NR & INV \\
        & Cell Function (2) & Tie-Low & XNR & OR & AOI & INV & XOR & AO & NAND & INV & NOR \\
        & \# Instances (1) & 29 & 34 & 10461 & 4829 & 29 & 9 & 8 & 1099 & 117 & 663 \\
        & \# Instances (2) & 2 & 1 & 26 & 564 & 17 & 108 & 4 & 827 & 17 & 46 \\
        & Similarity Score & 1.00 & 0.89 & 0.86 & 0.86 & 0.83 & 0.82 & 0.80 & 0.80 & 0.77 & 0.77 \\
        \midrule
        \multirow{5}{*}{\textbf{28\,nm}} 
        & Cell Function (1) & INV & INV & XNR & INV & INV & BUFF & INV & INV & BUFF & INV \\
        & Cell Function (2) & BUFF & BUFF & XOR & BUFF & BUFF & INV & BUFF & BUFF & INV & BUFF \\
        & \# Instances (1) & 212 & 167 & 348 & 462 & 1398 & 38 & 424 & 424 & 212 & 1398 \\
        & \# Instances (2) & 518 & 518 & 5379 & 1146 & 1 & 14 & 212 & 38 & 14 & 1146 \\
        & Similarity Score & 1.00 & 1.00 & 1.00 & 1.00 & 1.00 & 0.99 & 0.99 & 0.99 & 0.99 & 0.99 \\
        \bottomrule
    \end{tabular}
\end{table}
\smallskip

\subsection{Comparison of Our Detection Method with the Results of \texorpdfstring{Puschner~\etal}{Puschner et al}}
\label{appendix:comparison}
\leavevmode 

\begin{table}[H]
    \centering
    \caption{Detection results for the standard cell replacements from the original experiment of \citet{DBLP:conf/sp/PuschnerMBKMP23}. We compare the detection results of our novel metric with the detection results of Puschner~\etal}
    \begin{tabular}{c|ccc|ccc}
    \toprule
        & \multicolumn{3}{c|}{\textbf{Our Method}} & \multicolumn{3}{c}{\textbf{Method of Puschner~\etal}} \\
        Technology Node & True Positives & False Negatives & False Positives & True Positives & False Negatives & False Positives \\
        \midrule
        90 nm & 6 & 0 & 30 & 6 & 0 & 136 \\
        65 nm & 6 & 0 & $>$ 5000 & 6 & 0 & 6 \\
        40 nm & 6 & 0 & 16 & 6 & 0 & 11 \\
        28 nm & 6 & 0 & 91 & 3 & 3 & 343 \\
    \bottomrule
    \end{tabular}
    \label{tab:detectioncomparison}
\end{table}

\end{document}